\documentstyle[preprint,aps,epsf]{revtex}

\tighten
\draft

\begin{document}
\title{{\bf Liquid-gas phase transition in a two-components isospin lattice gas
model for asymmetric nuclear matter}}
\author{Wei Liang Qian$^{1,3}$ Ru-Keng Su$^{2.3}$}
\address{$^{1}$Surface Physics Lab (National Key Lab),\\
Fudan University, Shanghai 200433, P.R.China}
\address{$^{2}$China Center of Advanced Science and Technology (World Laboratory),\\
P. O. Box 8730, Beijing, P.R.China}
\address{$^{3}$Department of Physics, Fudan University, Shanghai 200433, P.R.China}
\maketitle

\begin{abstract}
A two-components isospin lattice gas model has been employed to study the
liquid-gas phase transition for asymmetric nuclear matter. An additional
degree of freedom, namely, the asymmetry parameter $\alpha $ has been
considered carefully for studying the phase transition. We have shown that
under the mean field approximation, the liquid-gas phase transition given by
this model is of second order. The entropy continues at the phase transition
point. The binodal surface is addressed.
\end{abstract}


An one-component lattice gas model with one type of atom has been found to
be successful for describing the liquid-gas phase transition (LGPT)\cite
{le52}. An isospin lattice gas model which assumes a lattice with each site
either being vacant or occupied by proton and neutron has been employed by
many authors to study the LGPT of {\it symmetric} nuclear matter recently
[2-6]. This model is mapped into a spin-1 Ising model. It has been proven
that the LGPT of this model is of first order under the Bragg-Williams mean
field approximation. The chemical potential continues at the phase
transition point but its first order derivatives, namely, entropy and
volume, are discontinuous. This result is reasonable if we notice that the
number of protons $N_{+}$ equals to the number of neutrons $N_{-}$ for
symmetric nuclear matter and the symmetric isospin lattice gas model is, in
fact, a system with one-component.

However, as was pointed out by ref.[7-12], the LGPT for a multi-components
system is of second order, i.e. the entropy continues and the second order
derivatives of chemical potential are discontinuous. This is because the
degrees of freedom increase according to the Gibbs phase rule, and then the
binodal surface has greater dimensionality. Since the space occupied by
nuclear matter addressed by ref.[7-12] is continuous, the above conclusion
is of course limited to a continuous manifold of space structure. It is of
interest to extend this study to a two-components system with lattice space
structure.

The isospin lattice gas model for asymmetric nuclear matter suggested by
ref.[13] is a typical two-components model because $N_{+}\neq N_{-}$, the
number of protons and neutrons are independent. We will employ this model to
reconsider this problem. In conflict with ref.[13], we will show the LGPT is
of second order for two-components isospin lattice gas model and the binodal
surface will end at two points with $\alpha =0$ and $\alpha =1$ where $%
\alpha $ is the asymmetric parameter 
\begin{equation}
\alpha =\frac{N_{-}-N_{+}}{N_{-}+N_{+}}
\end{equation}
At these two end-points the LGPT becomes first order because the nuclear
matter reduces to symmetric and neutron matter, and the lattice gas model
reduces to one-component.

The Hamiltonian of lattice gas model is[13]

\begin{equation}
H=-\sum_{<ij>}J_{ij}\tau _{zi}\tau _{zj}-h\sum_i\tau _{zi}  \label{Hint}
\end{equation}
where $<ij>$ denotes a nearest neighbor pair, $h$ is a constant and the
interaction strength parameter $J_{ij}$ satisfy 
\begin{equation}
J_{ij}=\left\{ 
\begin{array}{ll}
J_{ij} & {\normalsize for\ nearest\ neighbor\ distance\ }=a \\ 
\infty & {\normalsize for\ neighbor\ distance\ }=0 \\ 
0 & {\normalsize otherwise}
\end{array}
\right.  \label{pwell}
\end{equation}
The isospin $\tau _z$ is 
\begin{equation}
\tau _z=\left\{ 
\begin{array}{ll}
-1 & {\normalsize for\ neutron} \\ 
0 & {\normalsize for\ vacancy} \\ 
+1 & {\normalsize for\ proton}
\end{array}
\right.
\end{equation}
The Hamiltonian can be written as, 
\begin{equation}
H=-J_s(N_{++}+N_{--})+J_dN_{+-}-h\left( N_{+}-N_{-}\right)  \label{h2}
\end{equation}
where $N_{++}$, $N_{--}$ and $N_{+-}$...represent the nearest neighbor pairs
of proton-proton, neutron-neutron and proton-neutron respectively, and the
interaction strength parameter $J_s$ for $pp$ and $nn$ pairs, and $J_d$ for $%
pn$ pairs. Introducing 
\begin{equation}
r=\frac{N_0}N, 
\begin{array}{l}
\end{array}
s=\frac{N_{+}-N_{-}}N, 
\begin{array}{l}
\end{array}
N=N_{+}+N_{-}+N_0
\end{equation}
where $r$ denotes the relative emptiness and $s$ plays the similar role as
that of asymmetric parameter $\alpha $. $s$ is not a new parameter. We
introduce it for comparing with ref.[13] only. By using the Bragg-Williams
mean field approximation \cite{kh63} 
\begin{equation}
\frac{N_{++}}{\frac 12\gamma N}\approx \left( \frac{N_{+}}N\right) ^2,\frac{%
N_{--}}{\frac 12\gamma N}\approx \left( \frac{N_{-}}N\right) ^2,\frac{N_{00}%
}{\frac 12\gamma N}\approx \left( \frac{N_0}N\right) ^2  \label{BW2c}
\end{equation}
where $\gamma $ denotes the number of nearest neighbors of any given site,
and $\frac 12\gamma N$ is the total number of pairs. We can rewrite Eq.(\ref
{h2}) as 
\begin{equation}
H(r,s,N)=-C_1Ns^2-C_2N(1-r)^2-hNs
\end{equation}
where 
\begin{equation}
C_1=\frac{(J_s+J_d)\gamma }4,\;C_2=\frac{(J_s-J_d)\gamma }4
\end{equation}

The grand partition function of our system is 
\begin{equation}
Q_G=\sum_{r,s}g\left( r,s,N\right) z_{+}^{N_{+}}z_{-}^{N_{-}}e^{-\beta
H}=\sum_{r,s}g\left( r,s,N\right) z_{+}^{\frac N2\left( 1-r+s\right) }z_{-}^{%
\frac N2\left( 1-r-s\right) }e^{-\beta H}  \label{T2c}
\end{equation}
where $\beta =\frac 1{k_BT}$, the factor 
\begin{equation}
g\left( r,s,N\right) =\frac{N!}{N_0!N_{+}!N_{-}!}=\frac{N!}{\left( Nr\right)
!\left[ \frac N2\left( 1-r+s\right) \right] !\left[ \frac N2\left(
1-r-s\right) \right] !}  \label{g2c}
\end{equation}
the fugacity $z_{+}=e^{\beta \mu _{+}}$ and $z_{-}=e^{\beta \mu _{-}}$ for
proton and neutron respectively. For two-components system, in general, the
chemical potential of proton $\mu _{+}$ does not equal to the chemical
potential of neutron $\mu _{-}$ because $N_{+}\neq N_{-}$. According to the
Gibbs phase rule, the degrees of freedom of a system is 
\begin{equation}
f=k-\varphi +2
\end{equation}
where $k$ and $\varphi $ are the number of components and phases
respectively. A two-components system has one more degree of freedom than
that of one-component. We choose the asymmetric parameter $\alpha $ (or $s$)
as the additional degree of freedom for two-components lattice gas model as
that of nuclear matter models with continuous space structure [7-12].
Therefore, we have two independent parameters, namely $r$ (corresponding to
the density) and $s$ (corresponding to the asymmetry of protons and
neutrons) in a two-components lattice gas model for asymmetric nuclear
matter. This is the basical difference between our calculation and that of
ref.[13].

In the thermodynamic limit, the sum in Eq.(\ref{T2c}) can be replaced by its
most dominant term [3,13,14]. Since we have two independent parameters $r$
and $s$, the extreme values ($\bar{r}$, $\bar{s}$) satisfy 
\begin{equation}
\frac{\partial \ln Q_G}{\partial r}|_{r=\bar{r},s=\bar{s}}=0  \label{e1}
\end{equation}
and 
\begin{equation}
\frac{\partial \ln Q_G}{\partial s}|_{r=\bar{r},s=\bar{s}}=0  \label{e2}
\end{equation}
We must use two equations to determin the extreme values of $r$ and $s$.
Using stirling's formula and Eqs.(\ref{e1}) (\ref{e2}), we obtian 
\begin{equation}
\beta \mu _{+}=\log z_{+}=-\log \bar{r}+\log \frac{1-\bar{r}+\bar{s}}2%
-2\beta C_2\left( 1-\bar{r}\right) -2\beta C_1\bar{s}-\beta h  \label{root5}
\end{equation}
\begin{equation}
\beta \mu _{-}=\log z_{-}=-\log \bar{r}+\log \frac{1-\bar{r}-\bar{s}}2%
-2\beta C_2\left( 1-\bar{r}\right) +2\beta C_1\bar{s}+\beta h  \label{root6}
\end{equation}
Only the maximum of the extreme values gives us the most dominant
contribution of Eq.(\ref{T2c}) and corresponds to stable state. Taking $h=0$
to neglect the effect of the ''external field'', we find Eq.(\ref{root5})
reduces to Eq.(\ref{root6}) and $\mu _{+}=\mu _{-}$ when $\bar{s}=0$, the
asymmetric nuclear matter becomes symmetric one and two-components becomes
one-component. The symmetric nuclear matter is the specific case of our
theory.

By means of the grand partition function, we can obtain other
thermodynamical quantities such as pressure, baryon density and entropy per
baryon easily. They are 
\begin{eqnarray}
P_{gas}\left( \bar{r},\bar{s},\beta \right) &=&C_1\bar{s}^2+C_2(1-\bar{r})^2-%
\frac{(1-\bar{r}+\bar{s})}{2\beta }\log \frac{(1-\bar{r}+\bar{s})}2-\frac{(1-%
\bar{r}-\bar{s})}{2\beta }\log \frac{(1-\bar{r}-\bar{s})}2  \nonumber \\
&&-\frac{\bar{r}}\beta \log \bar{r}+h\bar{s}+\frac{\left( 1-\bar{r}+\bar{s}%
\right) }{2\beta }\log z_{+}+\frac{\left( 1-\bar{r}-\bar{s}\right) }{2\beta }%
\log z_{-}  \label{pressure}
\end{eqnarray}
\begin{equation}
\rho =1-\bar{r}  \label{rho}
\end{equation}

\begin{eqnarray}
\frac SB &=&\frac S{N_{+}+N_{-}}=\frac 1{1-\bar{r}}\frac SN  \nonumber \\
&=&-\frac 1{1-\bar{r}}\left( \frac{\left( 1-\bar{r}+\bar{s}\right) }2\log 
\frac{\left( 1-\bar{r}+\bar{s}\right) }2+\frac{\left( 1-\bar{r}-\bar{s}%
\right) }2\log \frac{\left( 1-\bar{r}-\bar{s}\right) }2+\bar{r}\log \bar{r}%
\right)  \label{entropy}
\end{eqnarray}

Now we are in the position to discuss the LGPT of our model. The two-phase
coexistence equations read 
\begin{equation}
\mu _{+}^L=\mu _{+}^V  \label{m1}
\end{equation}
\begin{equation}
p^L=p^V  \label{p}
\end{equation}
\begin{equation}
\mu _{-}^L=\mu _{-}^V  \label{m2}
\end{equation}
where subsripts of one phase $L$ and $V$ stand for liquid and gas
respectively. The parameters $C_1$, $C_2$ are determined as follows: to have
an attractive nearest neighbor interaction, we have $J_s>0$ and $J_d<0$. In
this case $C_2$ is positive. We choose the values of $C_1$ and $C_2$ to
reproduce the binding energy ($-16MeV$) for symmetric nuclear matter at
saturation point when $T=0K$, and find 
\[
C_1=3.2MeV, \ \ \ C_2=16.0MeV 
\]

Our numerical results are shown in figures(1-4). In Fig.1, we show the
isotherms for fixed $\alpha =0.50$ but different temperature $T=6.00,$ $%
6.90, $ $7.95,$ $8.58$ and $8.70MeV$ respectively. The stable states of
isotherms corresponding to the most dominant term in the summation of Eq.(%
\ref{T2c}) is drawn by the solid curves. The dotted curve which encloses the
transition region represents the slice of binodal surface with constant
asymmetric parameter. The critical temperature for $\alpha =0.50$ asymmetric
nuclear matter $T_c=8.58MeV$.

The chemical isobar of proton and neutron vs. $\alpha $ curves at fixed
temperature $T=7.50MeV$ and $p=1.12MeVfm^{-3}$ are shown in Fig.2. In this
figure, all metastable and unstable states are neglected. The Gibbs
conditions Eqs.(20-22) for phase transition demand equal pressures and equal
chemical potentials for two phase with different $\alpha $. The collection
of all such $\alpha _1\left( T,p\right) $ and $\alpha _2\left( T,p\right) $
form the binodal surface. The two desired solutions form the edges of a
rectangle and can be found by using the geometrical construction shown in
Fig.2. We find the end points of the phase boundary are just mapped at the
edges of the rectangle. This situation is the same as that of the models
with continuous space manifold [7-12].

The section of binodal surface at $T=7.50MeV$ are shown in Fig.3. We find
that the end-points of binodal surface are fixed at $\alpha =0$ and $\alpha
=1$ respectively. The pressure and chemical potential unchange during phase
transition. This is just the behavior of one-component system. For cases $%
0<\alpha <1$, the physical behavior of Fig.3 can be explained as follows:
assume that the system is initially prepared with $\alpha =0.5$ (gas phase),
during isothermal compression, the two phase coexistence region is
encountered at point $A$, and the liquid phase emerges at point $B$. The
isothermal compression process continues to the point $C$ of liquid-gas
coexistence region, the density $\rho _{+}$ and $\rho _{-}$ for protons and
neutrons satisfy 
\begin{eqnarray}
\rho _{+} &=&\lambda \rho _{+}^V+\left( 1-\lambda \right) \rho _{+}^L 
\nonumber \\
\rho _{-} &=&\lambda \rho _{-}^V+\left( 1-\lambda \right) \rho _{-}^L\ \ \ \
\ \ \left( 0<\lambda <1\right)  \label{cor}
\end{eqnarray}
where $\lambda $ refers to the volume fraction of gas phase, when the
process evolves to the point $F$, the gas phase disppears and the system
becoms liquid.

Finally, we investigate the entropy and the order of LGPT. The entropy per
baryon as a function of pressure for different asymmetric parameter $\alpha $
are shown in Fig.4. The states denoted by dotted curves in the intermedial
states during the phase transition is evaluated according to 
\begin{equation}
\frac SB=\lambda \frac{S^V\rho ^V}{B^V\rho }+\left( 1-\lambda \right) \frac{%
S^L\rho ^L}{B^L\rho }
\end{equation}
where $\rho =\rho _{+}+\rho _{-},\rho ^V=\rho _{+}^V+\rho _{-}^V,\rho
^L=\rho _{+}^L+\rho _{-}^L$ and the conservation law Eqs.(\ref{cor}). The
curves in Fig.4 show the order of phase transition transparently. For
symmetric nuclear matter ($\alpha =0$) and for neutron matter ($\alpha =1$),
the entropy discontinues at the phase-transitional point. The LGPT is of
first order for one-component system. For asymmetric nuclear matter, $%
0<\alpha <1$, the entropy continues at the phase-transitional point and the
LGPT is of second order.

In summary, contrary to ref.[13], an additional degree of freedom $\alpha $
which makes the chemical potential of proton does not equal to that of
neutron has been considered for asymmetric isospin lattice gas model. We
have shown that under Bragg-Williams mean field approximation, the LGPT in a
two-components isospin lattice gas model for asymmetric nuclear matter is of
second order. At the phase transition point, the entropy continues. The
binodal surface. The binodal surface for fixed temperature $T=7.50MeV$ has
been given.

\subsection{Acknowledgements}

This work was supported in part by the National Natural Science Fundation of
China under constract No. 19975010, 10047005, 19947001.

\begin{center}
{\bf {FIGURE\thinspace \thinspace \thinspace CAPTIONS}}
\end{center}

\begin{description}
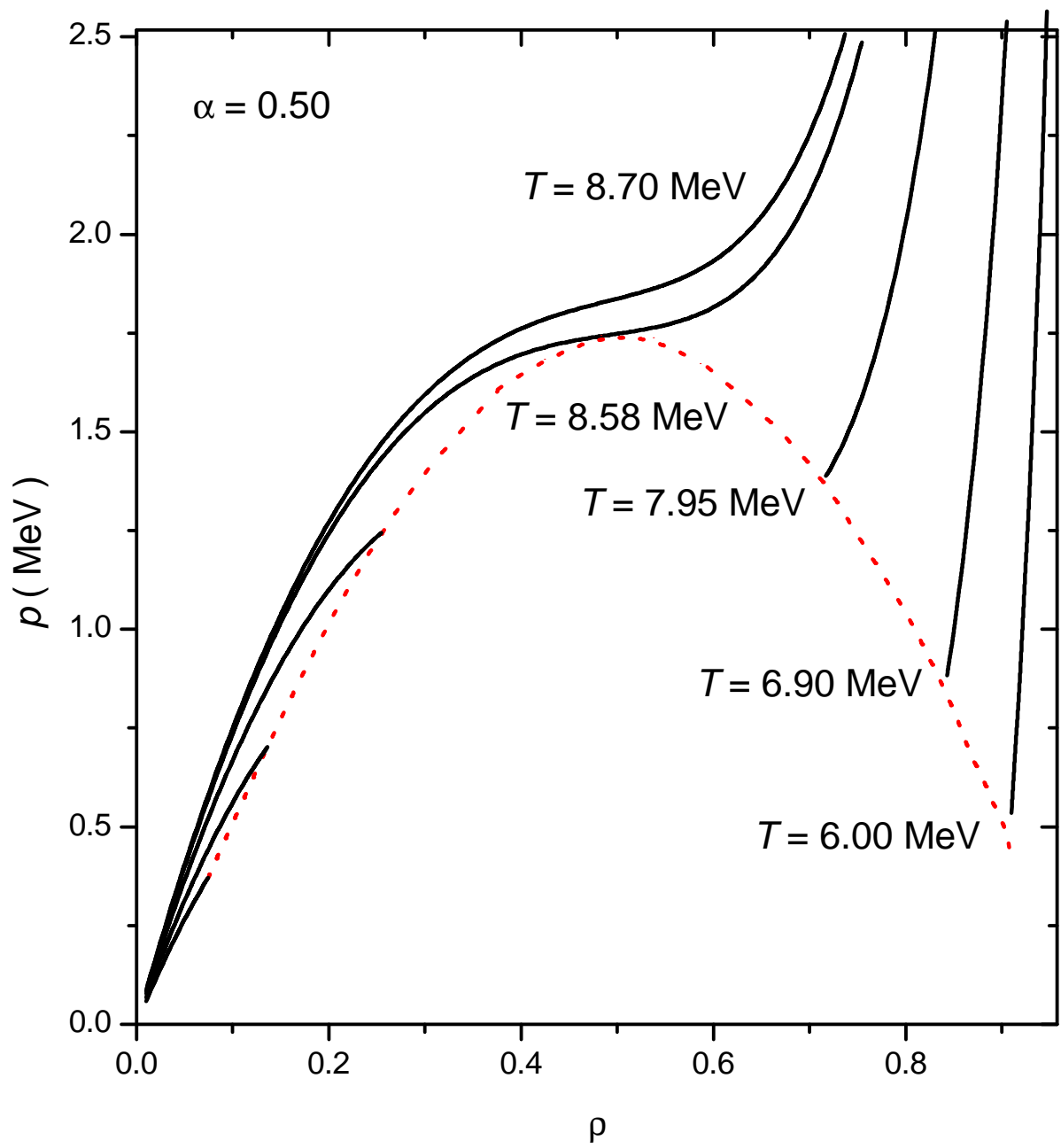
\item[Fig.1]  The isotherms for fixed $\alpha =0.50$ and different
temperatures $T=6.00,6.90,7.95,8.58$ and $8.70MeV$ respectively.

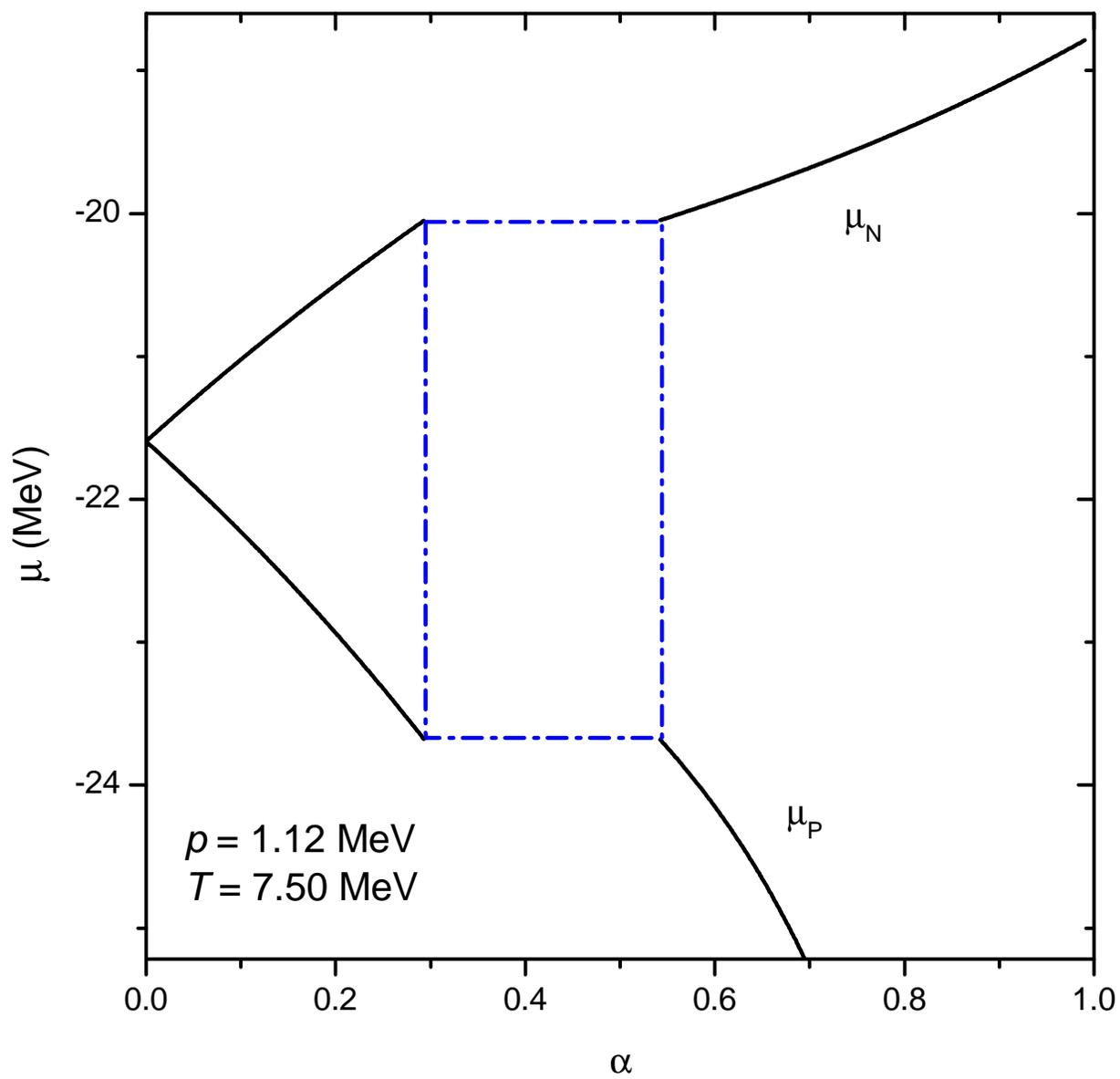
\item[Fig.2]  Chemical potentials for proton and neutron as a function of $%
\alpha $ at $p=1.12MeV$ and $T=7.50MeV$.

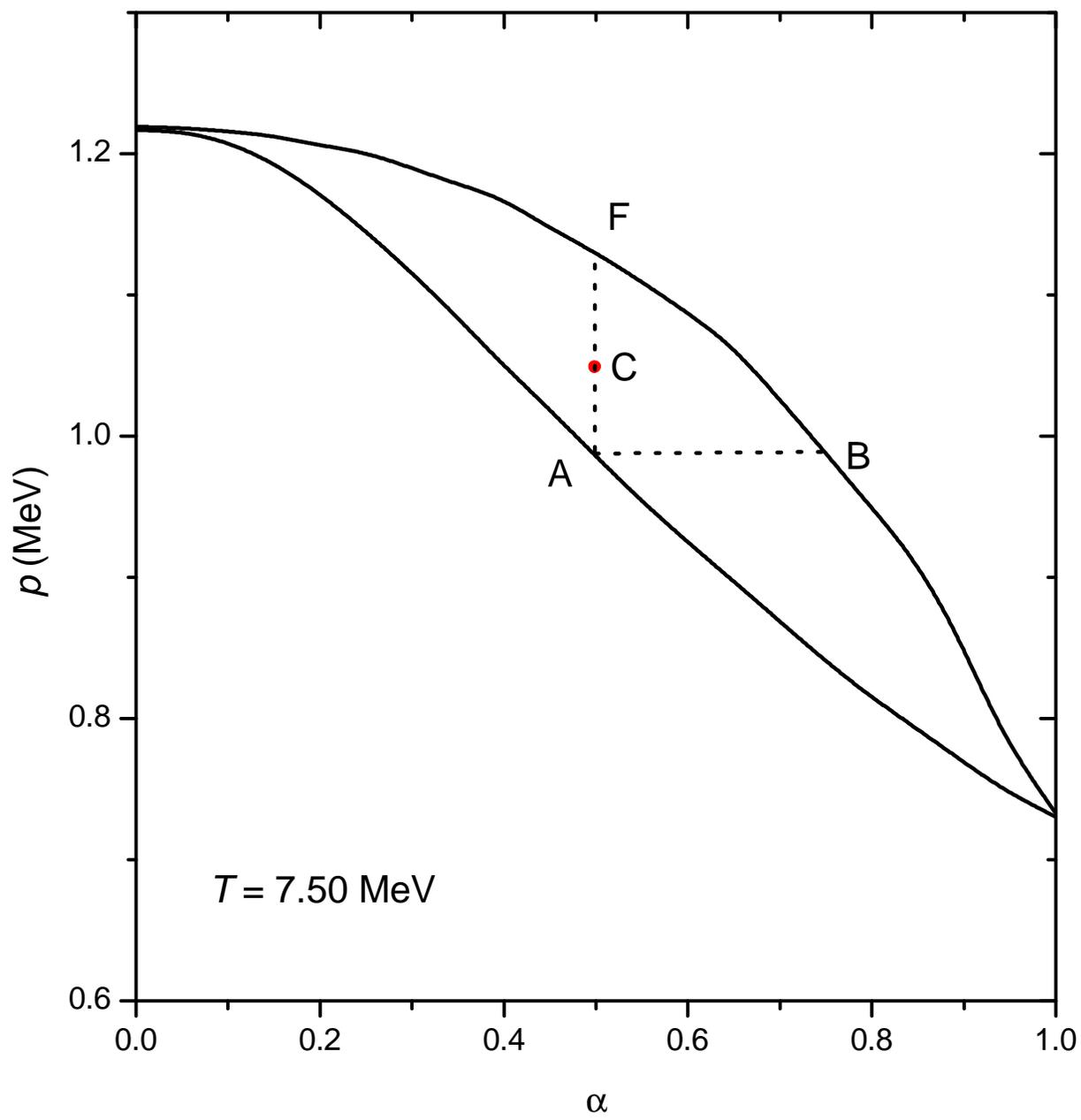
\item[Fig.3]  The section of binodal surface for $T=7.50MeV$.

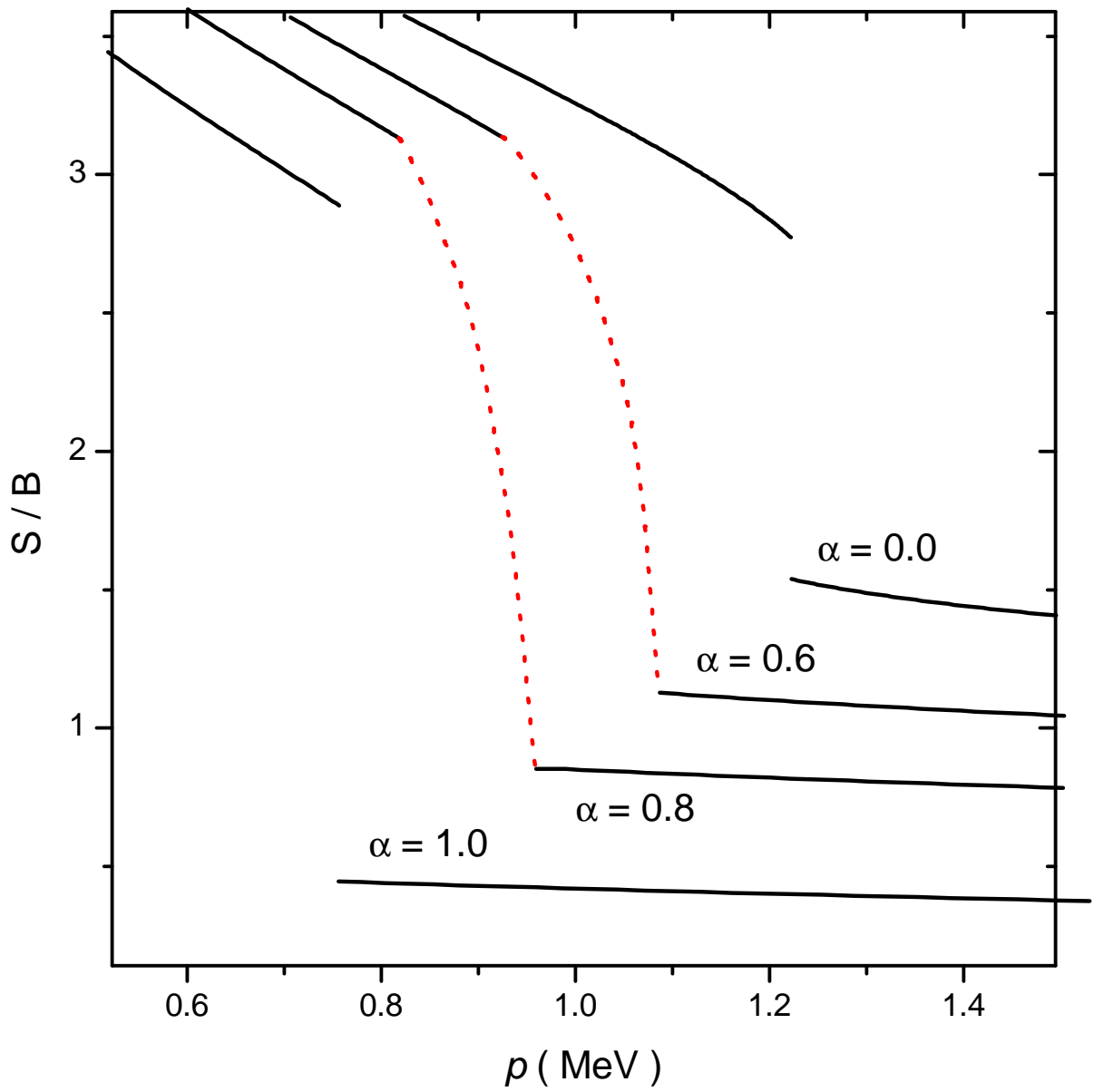
\item[Fig.4]  Entropy per baryon vs. pressure for asymmetric parameter $%
\alpha =0.0,\alpha =0.6,\alpha =0.8$ and $\alpha =1.0$ at $T=7.5MeV$.
\end{description}

\end{document}